\documentclass[amsmath,amssymb,pre]{revtex4}
\usepackage{bm}
\usepackage{graphicx}

\renewcommand{\vec}[1]{\bm{#1}}

\begin{document}
\title{Configurational Temperature in Membrane Simulations 
  Using Dissipative Particle Dynamics}%
\author{Michael P. Allen}
  \affiliation{Department of Physics and Centre for Scientific Computing,
  University of Warwick, Coventry CV4 7AL, United Kingdom}
\begin{abstract}
  The use of excessively long timesteps in dissipative particle dynamics
  simulations may produce simulation artifacts due to the generation of
  configurations which are not representative of the desired canonical ensemble.
  The configurational temperature, amongst other quantities, may be used to
  assess the extent of the deviation from equilibrium. This paper presents
  results for simulations of models of water, and lipid bilayer membranes, to
  illustrate the nature of the problems.
\end{abstract}
\maketitle
\section{Introduction}
Dissipative particle dynamics \citep{hoogerbrugge.pj:1992.a,koelman.jmva:1993.a}
(DPD) has become a popular tool for simulating the behaviour of both simple and
complex fluids. In outline, it consists of the solution of the classical
equations of motion for a system of interacting particles, together with a set
of stochastic and dissipative forces which control the temperature and allow one
to choose the viscosity. It is usual to employ a very simple, repulsive, pair
potential, often choosing the force law derived from it to vary linearly with
separation within a specified cutoff range. As well as these conservative
forces, the stochastic and dissipative forces also act in a pairwise fashion, so
as to conserve momentum and ensure hydrodynamic behaviour. The particles
represent fluid regions, rather than individual atoms and molecules: the
softness and simplicity of the interactions permit the use of a long time step,
compared with conventional molecular dynamics. This, and the acceleration of
physical processes compared with those seen in more realistic simulations, gives
an advantage of several orders of magnitude, at the cost of a very rough mapping
onto specific molecular properties. The way to relate the key DPD parameters to
fluid equations of state and transport coefficients, has been usefully presented
by \citet{groot.rd:1997.a}.

By tuning the interactions to represent hydrophilic and hydrophobic
units, and linking together the particles in a simple bead-spring
fashion, the DPD method has been applied to lipid bilayer
membranes \citep{groot.rd:2001.a}. Self-assembly and phase diagrams have been
investigated \citep{kranenburg.m:2003.b}, stress profiles
determined \citep{shillcock.jc:2002.a}, and the effects of small
molecules \citep{kranenburg.m:2004.b} and membrane-bound
proteins \citep{venturoli.m:2005.a} studied.

Several papers have attempted to optimise the computational algorithm used to
solve the DPD
equations \citep{pagonabarraga:1998.a,lowe.cp:1999.a,gibson.jb:1999.a,besold.g:2000.b,%
denotter.wk:2001.a,vattulainen.i:2002.a,nikunen.p:2003.a,shardlow.t:2003.a,%
peters.eajf:2004.a,stoyanov.sd:2005.a}. %
The early emphasis was on the correct handling of the velocity-dependent
dissipative forces, and ensuring the correct balance of these with the random
forces, at an appropriate timestep, so as to generate the canonical
ensemble \citep{espanol.p:1995.b}. Accordingly, most
attention \citep{groot.rd:1997.a} has focused on maintaining a kinetic temperature
$T_\text{k}$ (defined in terms of the kinetic energy per degree of freedom, via
the equipartition theorem) reasonably close to the desired temperature $T$. A
detailed analysis of the ideal fluid limit \citep{marsh.ca:1997.a}, as a function
of timestep, has been carried out. In addition, it has become standard practice
to check the pair distribution function and diffusion coefficient for dependence
on the time step, as diagnostic tests.

Recently, increased attention has been paid to the suitability or otherwise of
the time step for the integration of the conservative forces. This is
particularly important in simulations of multi-component systems, polymers, and
membranes, where some of the interaction potentials may be stronger than others,
including intramolecular bond-bending and stretching terms. The usual criterion
of molecular dynamics, energy conservation in the absence of random and
dissipative forces, is not usually employed, and any problems may be disguised
by the thermostatting. In order to generate meaningful results, it is essential
to sample configurations from the canonical (Boltzmann) distribution, and any
indicator of significant deviations must be taken seriously.
\citet{denotter.wk:2000.c} were the first to point out that DPD, at the usual
timesteps and state points, generates configurations for which the
\emph{configurational temperature} $T_\text{c}$ differs significantly from the
desired temperature $T$ and the kinetic temperature $T_\text{k}$.
\citeauthor{denotter.wk:2000.c} estimated this quantity by calculating (for
small timesteps) the average potential energy $\langle U\rangle$ as a function
of temperature in the vicinity of $T$, and fitting the results to a quadratic
curve; then, this curve was used to convert averages of $\langle U\rangle$
measured at larger timesteps into an estimate of $T_\text{c}$. Here a different,
and more straightforward definition is used.  The configurational temperature
$T_\text{c}$ is written
\begin{equation}
k_\text{B}T_\text{c} = \frac{\sum_j%
\left\langle \bigl| \vec{\nabla}_j U \bigr|^2 \right\rangle}%
{\sum_j\left\langle\nabla_j^2 U\right\rangle} \;,
\label{eqn:contemp}
\end{equation}
where $k_\text{B}$ is Boltzmann's constant, $\vec{\nabla}_j$ the gradient, and
$\nabla_j^2$ the laplacian, of the potential energy $U$, with respect to the
position of particle $j$. The summations may be over all particles in the
system, or restricted to a single species, or indeed to an individual particle.
The expression has long been known as a hypervirial relation
\citep{hirschfelder.jo:1960.a,powles.jg:2005.a}.  It was first introduced by
\citet{rugh.hh:1997.a} as an independent estimate of temperature in
simulations, and recommended as a diagnostic test for lack of equilibrium by
\citet{butler.bd:1998.a}. 

In the context of membrane simulations, \citet{jakobsen.af:2005.a} observed
significant differences in the kinetic temperatures of different species in DPD
simulations, and recommended this as a simple diagnostic test of incorrect
integration of the equations of motion.  They also mentioned the possibility of
monitoring the configurational temperature, but did not report any results for
this. However, they observed several other simulation artifacts, including
time-step dependence of the pressure profiles and particle densities within the
membrane.  \citet{hafskjold.b:2004.a} have reemphasized
that the timesteps typically recommended and used in DPD are too long to
properly handle the discontinuity at the cutoff of the usual conservative force
law: they investigated the possibility of softening this cutoff via spline
functions, but no dramatic improvements at long time step were seen, and they
recommended the use of shorter time steps.

This paper reports simulation averages of the configurational temperature for a
standard single-component DPD fluid, over a range of timesteps, using a variety
of proposed DPD integration methods. The results show that, although some of
these methods dramatically improve the control of the kinetic temperature
$T_\text{k}$, the same improvements are not seen for the configurational
temperature: deviations of $T_\text{c}$ from the desired values
remain significant in all cases, at the typically recommended time steps.
Results are then reported for the same membrane model used in
Ref.~\citep{jakobsen.af:2005.a}: largely as a result of intramolecular bonding
terms in the potential, the deviations in $T_\text{c}$ are far worse than in the
simple fluid case, and may be correlated with some of the other artifacts
observed before \citep{jakobsen.af:2005.a}. The general conclusion is that the
configurational temperature should be added to the list of diagnostic tests
applied to DPD simulations, and that shorter timesteps should be employed in
membrane simulations than hitherto.
\section{Dynamical Algorithms}
The DPD equations of motion for a simple fluid may be
written \citep{hoogerbrugge.pj:1992.a,koelman.jmva:1993.a,espanol.p:1995.b}
\begin{subequations}
\label{eqn:dpd}
\begin{align}
\text{d}\vec{r}_i &= \bigl(\vec{p}_i/m\bigr)\text{d}t 
\label{eqn:dpda}\\
\text{d}\vec{p}_i &= \sum_{j\neq i} \vec{f}_{ij}^\text{C}\text{d}t 
+ \vec{f}_{ij}^\text{D}\text{d}t 
+ \text{d}\vec{p}_{ij}^\text{R} \;.
\label{eqn:dpdb}
\end{align}
\end{subequations}
The particles are all assumed to have the same mass $m$.
The conservative forces usually take the form
\begin{subequations}
\begin{align}
 \vec{f}_{ij}^\text{C} &= \alpha \omega(r_{ij})\hat{\vec{r}}_{ij}
\label{eqn:conforcea}
\\
\text{with}\qquad
\omega(r) &= \begin{cases} 1-r/r_\text{c} & r\leq r_\text{c} \\ 0 & r>
  r_\text{c} \end{cases} \;.
\label{eqn:conforceb}
\end{align}
\end{subequations}
Here $\vec{r}_{ij}=\vec{r}_{i}-\vec{r}_{j}$, $r_{ij} = |\vec{r}_{ij}|$,
$\hat{\vec{r}}_{ij} = \vec{r}_{ij}/r_{ij}$. The parameter $\alpha$ determines
the strength of the conservative interactions, and $r_\text{c}$ is the cutoff.
The dissipative forces and random impulses are written
\begin{align*}
 \vec{f}_{ij}^\text{D} 
&= -\gamma \omega(r_{ij})^2 (\vec{v}_{ij}\cdot\hat{\vec{r}}_{ij})
\hat{\vec{r}}_{ij}
\\
\text{d}\vec{p}_{ij}^\text{R} &= \sigma \omega(r_{ij}) \hat{\vec{r}}_{ij}
\text{d}W_{ij} \;,
\end{align*}
where $\vec{v}_{ij}=\vec{v}_{i}-\vec{v}_{j}$, and $\vec{p}_i=m\vec{v}_i$.  A
choice has been made here to use the same weighting function $\omega(r)$ in the
specification of conservative, dissipative, and random terms.  The dissipative
friction $\gamma$ is related to the random impulse strength $\sigma$ by the
fluctuation-dissipation theorem $\sigma^2=2\gamma k_\text{B}T$.  The quantity
$\text{d}W_{ij}$ is the time derivative of a Wiener process: over a timestep
$\Delta t$, $\Delta W_{ij}$ is chosen from a normal distribution with zero mean
and variance $\Delta t$, and $\Delta W_{ji}=-\Delta W_{ij}$.  Units are chosen
such that the particle mass $m$, cutoff $r_\text{c}$, and temperature
$T$ are all unity.

Four algorithms will be discussed in this paper. The ``standard'' DPD algorithm
consists of the following steps \citep{gibson.jb:1999.a,vattulainen.i:2002.a}
\begin{subequations}
\begin{align}
\tilde{\vec{p}}_i &:= \vec{p}_i + \lambda \sum_{j\neq i}
\vec{f}_{ij}^\text{C}\Delta t 
+ \vec{f}_{ij}^\text{D}\Delta t 
+ \Delta\vec{p}_{ij}^\text{R} 
\label{alg:1a}\\
\vec{p}_i &:= \vec{p}_i + \frac{1}{2} \sum_{j\neq i} 
\vec{f}_{ij}^\text{C}\Delta t 
+ \vec{f}_{ij}^\text{D}\Delta t 
+ \Delta\vec{p}_{ij}^\text{R} 
\label{alg:1b}\\
\vec{r}_i &:= \vec{r}_i + \bigl(\vec{p}_i/m\bigr)\Delta t 
\label{alg:1c}\\
&\text{Calculate $\vec{f}_{ij}^\text{C}\bigl(\{\vec{r}_i\}\bigr)$, 
$\vec{f}_{ij}^\text{D}\bigl(\{\vec{r}_i\},\{\tilde{\vec{p}}_i\}\bigr)$,
$\Delta\vec{p}_{ij}^\text{R}\bigl(\{\vec{r}_i\}\bigr)$ }
\label{alg:1d}\\
\vec{p}_i &:= \vec{p}_i + \frac{1}{2} \sum_{j\neq i} 
\vec{f}_{ij}^\text{C}\Delta t 
+ \vec{f}_{ij}^\text{D}\Delta t 
+ \Delta\vec{p}_{ij}^\text{R} 
\label{alg:1e}\\
&\text{Calculate $\vec{f}_{ij}^\text{D}\bigl(\{\vec{r}_i\},\{\vec{p}_i\}\bigr)$}
\label{alg:1f}
\end{align}
\end{subequations}
Note the auxiliary momenta $\tilde{\vec{p}}_i$, used in the calculation of
dissipative forces, and predicted using a parameter $\lambda$. The value
$\lambda=0.5$ conveniently makes $\tilde{\vec{p}}_i= \vec{p}_i$, but
$\lambda=0.65$ was found by \citet{groot.rd:1997.a} to be optimal
for the particular fluid that will be studied here. This paper will use
$\lambda=0.65$.  Note also the second computation of dissipative forces, step
\eqref{alg:1f}, introduced by \citet{gibson.jb:1999.a}.
Iteration of the last two steps would lead to a self-consistent
algorithm \citep{besold.g:2000.b} not studied here.

The second algorithm is the so-called ``Scheme II'' of
\citet{peters.eajf:2004.a}. At each time step, the positions and momenta
are updated using the conservative forces only:
\begin{subequations}
\begin{align}
\vec{p}_i &:= \vec{p}_i + \frac{1}{2} \sum_{j\neq i} 
\vec{f}_{ij}^\text{C}\Delta t 
\label{alg:2a}\\
\vec{r}_i &:= \vec{r}_i + \bigl(\vec{p}_i/m\bigr)\Delta t 
\label{alg:2b}\\
&\text{Calculate $\vec{f}_{ij}^\text{C}\bigl(\{\vec{r}_i\}\bigr)$}
\label{alg:2c}\\
\vec{p}_i &:= \vec{p}_i + \frac{1}{2} \sum_{j\neq i} 
\vec{f}_{ij}^\text{C}\Delta t \;.
\label{alg:2d}
\end{align}
\end{subequations}
Following this, for every pair $i,j$, selected
in random order, the following update is performed:
\begin{align*}
\vec{p}_i &:= \vec{p}_i + \vec{f}_{ij}^\text{D}\Delta t +
\Delta\vec{p}_{ij}^\text{R} 
\\
\vec{p}_j &:= \vec{p}_j - \vec{f}_{ij}^\text{D}\Delta t -
\Delta\vec{p}_{ij}^\text{R} 
\\
\text{with}\qquad 
\vec{f}_{ij}^\text{D} &= -a_{ij} (\vec{v}_{ij}\cdot\hat{\vec{r}}_{ij}) \hat{\vec{r}}_{ij}
\\
\text{and}\qquad
\Delta\vec{p}_{ij}^\text{R} &= b_{ij} \Delta W_{ij} \hat{\vec{r}}_{ij} \;.
\end{align*}
The parameters $a_{ij}$ and $b_{ij}$ are determined by $T$ and $\gamma$ (or,
equivalently, $T$ and $\sigma$) and by the timestep. As explained by
\citet{peters.eajf:2004.a}, this generates the Maxwell-Boltzmann
distribution of velocities and, at short time step, solves the original DPD equations.

The third algorithm is due to \citet{lowe.cp:1999.a}. This is not a solution
of the original DPD equations, but rather a pairwise thermostat applied to a
conventional integration algorithm for the conservative forces. As for the
Peters method, at each step the positions and momenta are updated using steps
\eqref{alg:2a}--\eqref{alg:2d} above, using the conservative forces only. Then,
examining every pair (in random order), with probability $P=\Gamma\Delta t$ the
momenta are updated as follows:
\begin{align*}
\vec{p}_i &:= \vec{p}_i + \Delta\vec{p}_{ij}^\text{R} 
\\
\vec{p}_j &:= \vec{p}_j - \Delta\vec{p}_{ij}^\text{R} 
\\
\text{with}\qquad 
\Delta\vec{p}_{ij}^\text{R} &= \frac{1}{2}m\left[
\zeta \sqrt{2k_\text{B}T/m}
- (\vec{v}_{ij}\cdot\hat{\vec{r}}_{ij}) \right]\hat{\vec{r}}_{ij} \;,
\end{align*}
where $\zeta$ is selected from a Gaussian distribution with zero mean and unit
variance. This procedure reselects the relative velocity from the
Maxwell-Boltzmann distribution. The key parameter is the stochastic randomization
frequency $\Gamma$.

The fourth algorithm to be considered here has been proposed by
\citet{stoyanov.sd:2005.a}. It is identical with the method of Lowe, except that
the fraction $(1-P)$ of pairs which do not have their relative velocities
stochastically reselected, are instead thermalized by a deterministic method.
For each such pair, after steps \eqref{alg:2a}, \eqref{alg:2b}, a dissipative
force is calculated
\begin{align*}
\vec{f}_{ij}^\text{D} &= k \omega(r_{ij}) 
(\vec{v}_{ij}\cdot\hat{\vec{r}}_{ij})\hat{\vec{r}}_{ij} \\
\vec{f}_{i}^\text{D} &:=\vec{f}_{i}^\text{D}+\vec{f}_{ij}^\text{D} \\
\vec{f}_{j}^\text{D} &:=\vec{f}_{j}^\text{D}-\vec{f}_{ij}^\text{D} \;.
\end{align*}
Then, after steps \eqref{alg:2c}, \eqref{alg:2d}, the momenta are corrected by
\begin{align*}
\vec{p}_i &:= \vec{p}_i + \Delta t \bigl(1-\tilde{T}_\text{k}/T\bigr) \vec{f}_{i}^\text{D} \;,
\end{align*}
where $\tilde{T}_\text{k}$ is a kinetic temperature estimated from relative
velocities at the start of the timestep. Finally, the Lowe velocity reselection
process is applied to the remaining fraction $P$ of pairs as usual. In the
present work, following Ref.~\citep{stoyanov.sd:2005.a}, the parameter $k$ is
set to $k=0.3/\Delta t$. As for the Lowe method, the key parameter is the
stochastic randomization rate $\Gamma$ which determines $P$.
\citep{stoyanov.sd:2005.a} associate the deterministic part of their thermostat
with the Nos\'{e}-Hoover algorithm \citep{nose.s:1984.a,hoover.wg:1985.a}, but in
practice it seems to be more closely related to that of
\citet{berendsen.hjc:1984.a}, since the ``friction coefficient'' is not
a dynamical variable but is instead directly proportional to the temperature
factor $(1-\tilde{T}_\text{k}/T)$.  This approach permits a high degree of
temperature control at low, or even zero, values of $\Gamma$, where the Lowe
method is ineffective: indeed, \citeauthor{stoyanov.sd:2005.a} report an order of magnitude
improvement in $T_\text{k}$. However, the links with a statistical ensemble, and
with the DPD equations themselves, remain to be established.

\section{Simulation Models and Parameters}
By considering the equation of state and dynamical properties of water under
ambient conditions, \citet{groot.rd:1997.a} established the
standard parameter set that is commonly used in algorithm
tests \citep{vattulainen.i:2002.a,nikunen.p:2003.a,peters.eajf:2004.a,stoyanov.sd:2005.a}:
strength parameter $\alpha=25$, number density $\rho=3$, stochastic impulse
strength $\sigma=3$, and hence (for $T=1$) $\gamma=4.5$.  This is sometimes
referred to as ``Model B''. This paper reports results for Model B, and also for
variants with $\sigma=1$ and $\sigma=6$, using both the DPD and Peters methods.
For the Stoyanov-Groot and Lowe algorithms, stochastic collision rates
$\Gamma=0,0.2,4$ and the maximum possible $\Gamma=1/\Delta t$ (corresponding to
$P=1$, completely reselecting all velocities every timestep) are investigated. A
system size of $N=250$ particles in a cubic box with periodic boundary
conditions is used throughout.

For their tests of DPD simulations of membranes, %
\citet{jakobsen.af:2005.a} adopted the model of \citet{shillcock.jc:2002.a}.
Each lipid molecule has the form of a 7-bead chain $\text{H}\text{T}_6$ in which
repulsion parameters between hydrophilic ``head'' beads (H), hydrophobic
``tail'' beads (T), and ``water'' beads (W) are as follows:
$\alpha_\text{WW}=\alpha_\text{HH}=\alpha_\text{TT}=25$, $\alpha_\text{HW}=35$,
$\alpha_\text{HT}=50$, $\alpha_\text{TW}=75$. Successive lipid beads are
connected by harmonic springs, and a simple bond-bending potential favours
linearity:
\begin{subequations}
\begin{align}
 u_\text{stretch}(\vec{r}_{ij}) &= \tfrac{1}{2}k_\text{stretch}\bigl(r_{ij}-\ell_0\bigr)^2 \qquad
\text{with $k_\text{stretch}=128$, $\ell_0=0.5$} 
\label{eqn:potstretch}
\\
 u_\text{bend}(\vec{r}_{ij},\vec{r}_{jk}) 
&= k_\text{bend}(1-\hat{\vec{r}}_{ij}\cdot\hat{\vec{r}}_{jk})
\qquad \text{with $k_\text{bend}=20$} \;.
\label{eqn:potbend}
\end{align}
\end{subequations}
The present paper uses this model but follows \citet{jakobsen.af:2005.a} in
choosing all stochastic impulse strengths to be the same, $\sigma=3$, in the DPD
and Peters algorithms. For comparison, results obtained using the Stoyanov-Groot
method with $\Gamma=0$, and using the Lowe thermostat with $\Gamma=1/\Delta t$,
are reported.  A smaller system is used here than in the previous
papers \citep{shillcock.jc:2002.a,jakobsen.af:2005.a}: $N_\text{lipid}=100$ lipid
chains plus $N_\text{water}=2500$ water beads, making $N=3200$ beads in all. A
cuboidal box, with periodic boundaries, was employed.  Transverse box dimensions
were fixed so as to give an overall number density $\rho=3$, and an area per
lipid $2A/N_\text{lipid}=0.617$, close to the state of zero surface tension.
The chains were pre-formed into a flat bilayer, and allowed to equilibrate
before any measurements were made.

The time unit $\tau=r_\text{c}\sqrt{m/k_\text{B}T}$ is defined in terms of the
other fixed parameters. In these units, for the water model described above,
\citet{groot.rd:1997.a} recommend timesteps in the range $\Delta
t=0.04 \ldots 0.06$, so as to achieve an accuracy of order 1\% in $T_\text{k}$;
the present paper investigates time steps in the range $0.005 \leq \Delta t \leq
0.06$. Total simulation times for the water runs were $t_\text{run}=2000$, and
for the membrane runs were $t_\text{run}=10000$. Formulae for the Laplacians
used to calculate the configurational temperature for this model are given in
the Appendix.
\section{Results}

\begin{figure}[tbp]
\hrulefill
\caption{\label{fig:1}%
  Water simulations using (a) DPD algorithm with $\lambda=0.65$, (b) Peters
  scheme II.  Temperatures as functions of timestep $\Delta t$. Solid lines:
  configurational temperature $T_\text{c}$. Dashed lines: kinetic temperature
  $T_\text{k}$.  Circles: $\sigma = 1$. Squares: $\sigma=3$. Diamonds:
  $\sigma=6$. The lines are to guide the eye. Statistical errors are
  approximately the same size as the plotting symbols.}
\centerline{\includegraphics[width=0.6\textwidth,clip]{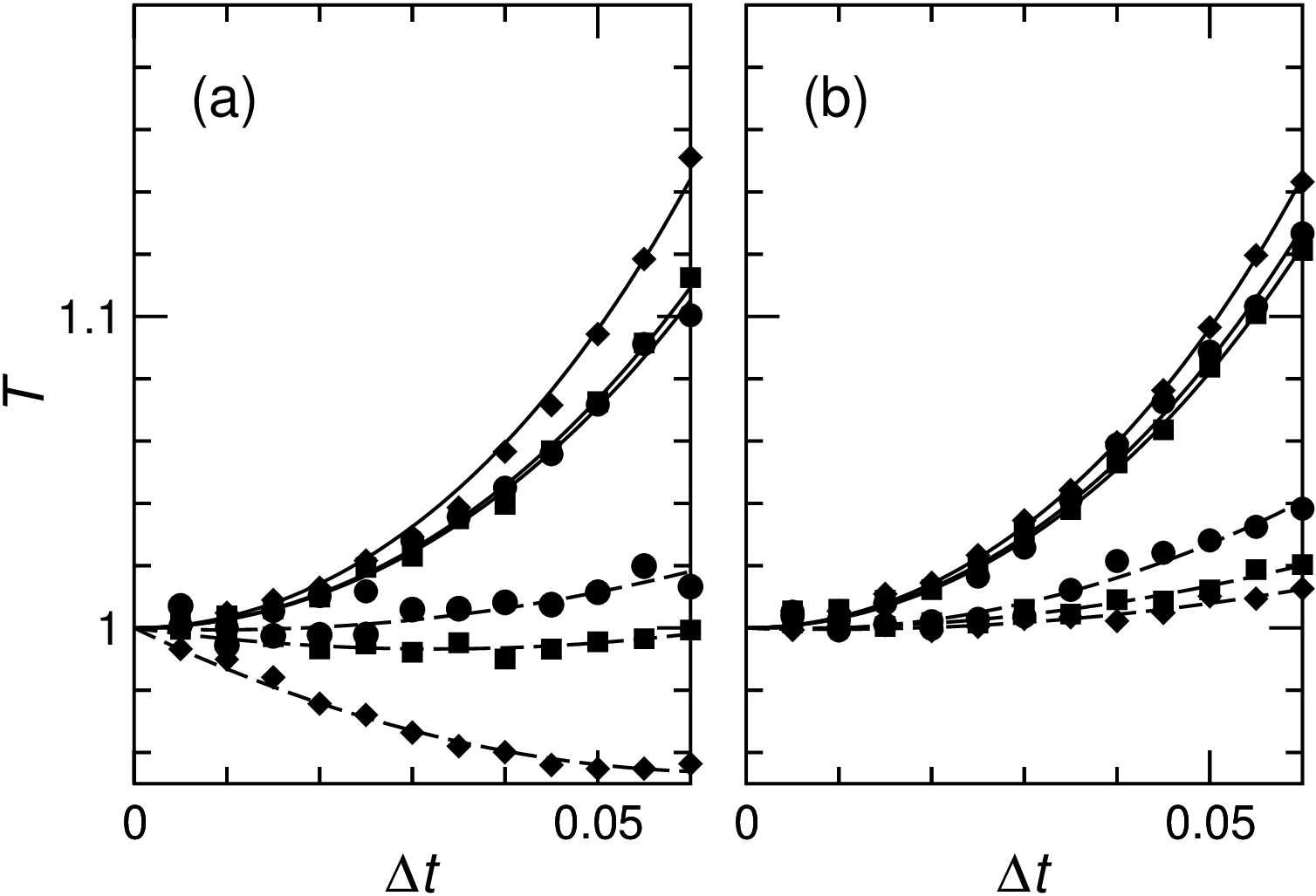}} \hrulefill
\end{figure}
\begin{figure}[tbp]
\hrulefill
\caption{\label{fig:2}%
  Water simulations using (a) Stoyanov-Groot thermostat, (b) Lowe thermostat.
  Temperatures as functions of timestep $\Delta t$.
  Solid lines: configurational temperature $T_\text{c}$. Dashed
  lines: kinetic temperature $T_\text{k}$.
  Downward-triangles: $\Gamma = 0$. Circles: $\Gamma=0.5$. Squares:
  $\Gamma=4$.  Upward-triangles: $\Gamma=1/\Delta t$ (maximum possible). The
  lines are to guide the eye. Statistical errors are approximately the same size
  as the plotting symbols.}
\centerline{\includegraphics[width=0.6\textwidth,clip]{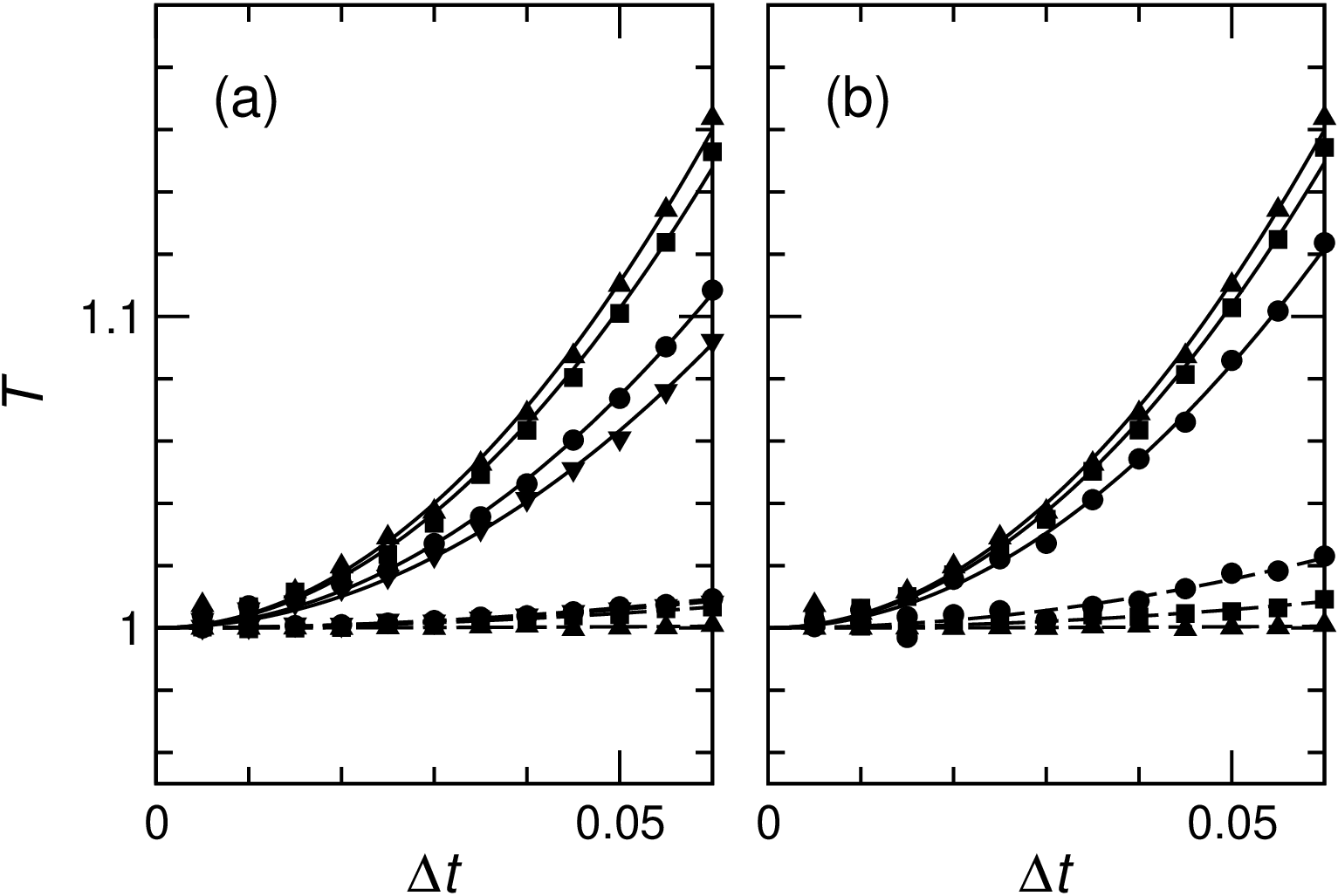}} \hrulefill
\end{figure}

Figures \ref{fig:1} and \ref{fig:2} show the measured temperatures for the four
chosen algorithms, in simulating the water model at various choices of
thermostatting strength.  With an appropriate choice of parameters, it is indeed
possible to achieve a kinetic temperature $T_\text{k}$ within 1\% of the desired
value at a timestep $\Delta t = 0.05$; the Stoyanov-Groot method performs well
in this regard at all values of $\Gamma$, and the Lowe thermostat is good at
high values of $\Gamma$.  The more recently-developed thermostats all seem less
sensitive to variations in $\sigma$ or $\Gamma$ than the basic DPD algorithm.
However, it can be seen that the error in the configurational temperature
$T_\text{c}$ is an order of magnitude worse than that in $T_\text{k}$, being
around 10\% at $\Delta t = 0.05$, and this is not dramatically affected by the
choice of algorithm: the ``best'' (Stoyanov-Groot at $\Gamma=0$) reduces the
error by 30--40\% compared to the ``worst'' (all of the methods, at high
stochastic thermalization rate). Also, for quite small stochastic damping
$\Gamma=0.5$ (i.e.\ $P=0.025$ at $\Delta t=0.05$), the Stoyanov-Groot method
becomes similar in performance to the other methods.

\begin{figure}[tbp]
\hrulefill
\caption{\label{fig:3}%
  Membrane simulations using (a) DPD algorithm with $\lambda=0.65$,
  $\sigma=3.0$, (b) Peters scheme II with $\sigma=3.0$.  Temperatures as
  functions of timestep $\Delta t$. Circles: H (head) and
  $\text{T}_6$ (tail) beads. Squares: $\text{T}_1$ and $\text{T}_5$ beads.
  Diamonds: $\text{T}_2$, $\text{T}_3$ and $\text{T}_4$ beads. Triangles: W
  (water) beads. The lines are to guide the eye. Statistical errors are
  approximately the same size as the plotting symbols.}
\centerline{\includegraphics[width=0.6\textwidth,clip]{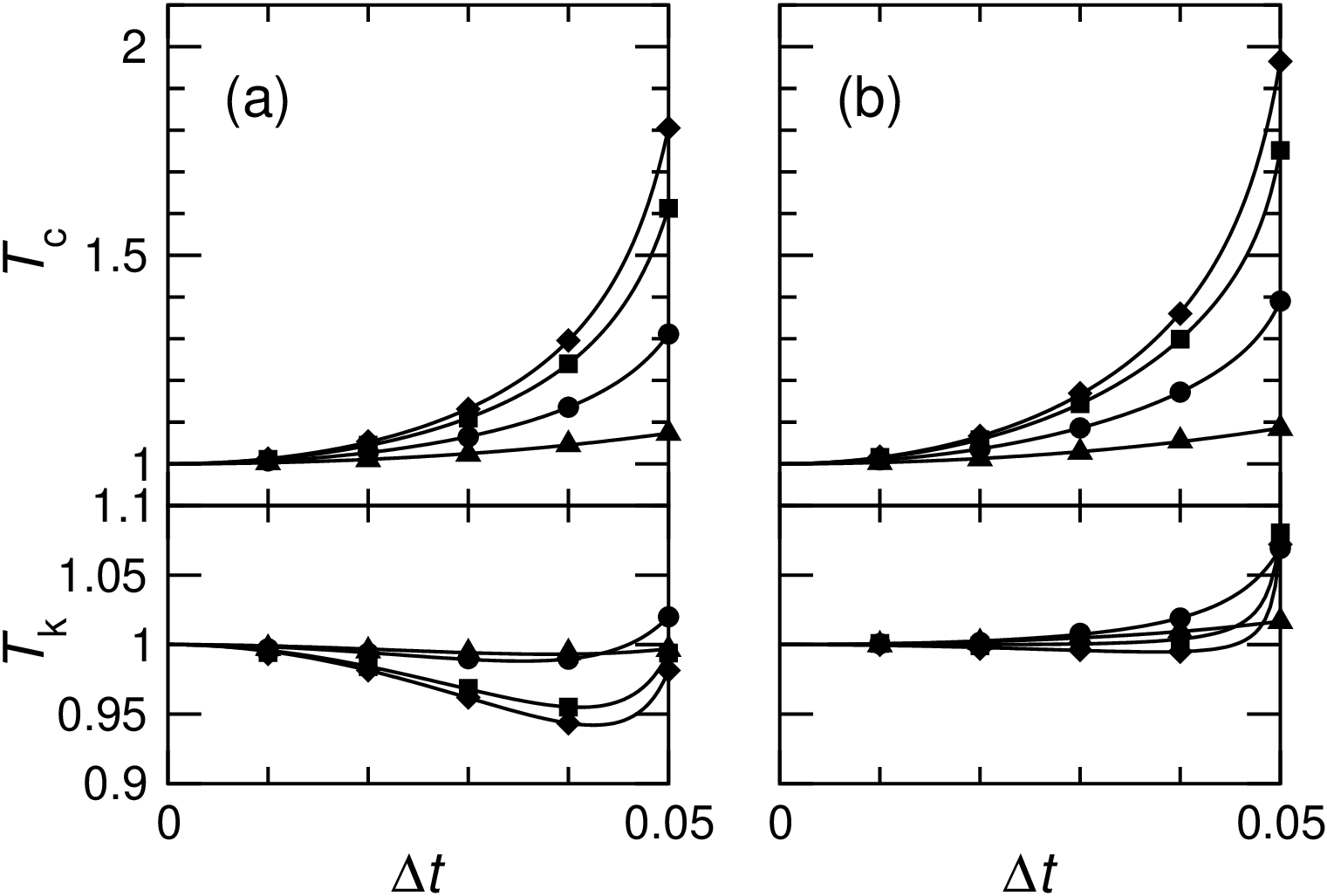}} \hrulefill
\end{figure}
\begin{figure}[tbp]
\hrulefill
\caption{\label{fig:4}%
  Membrane simulations using (a) Stoyanov-Groot thermostat with $\Gamma=0$, (b)
  Lowe thermostat with $\Gamma=1/\Delta t$.  Temperatures as functions of
  timestep $\Delta t$. Symbols as for Fig.~\ref{fig:3}.}
\centerline{\includegraphics[width=0.6\textwidth,clip]{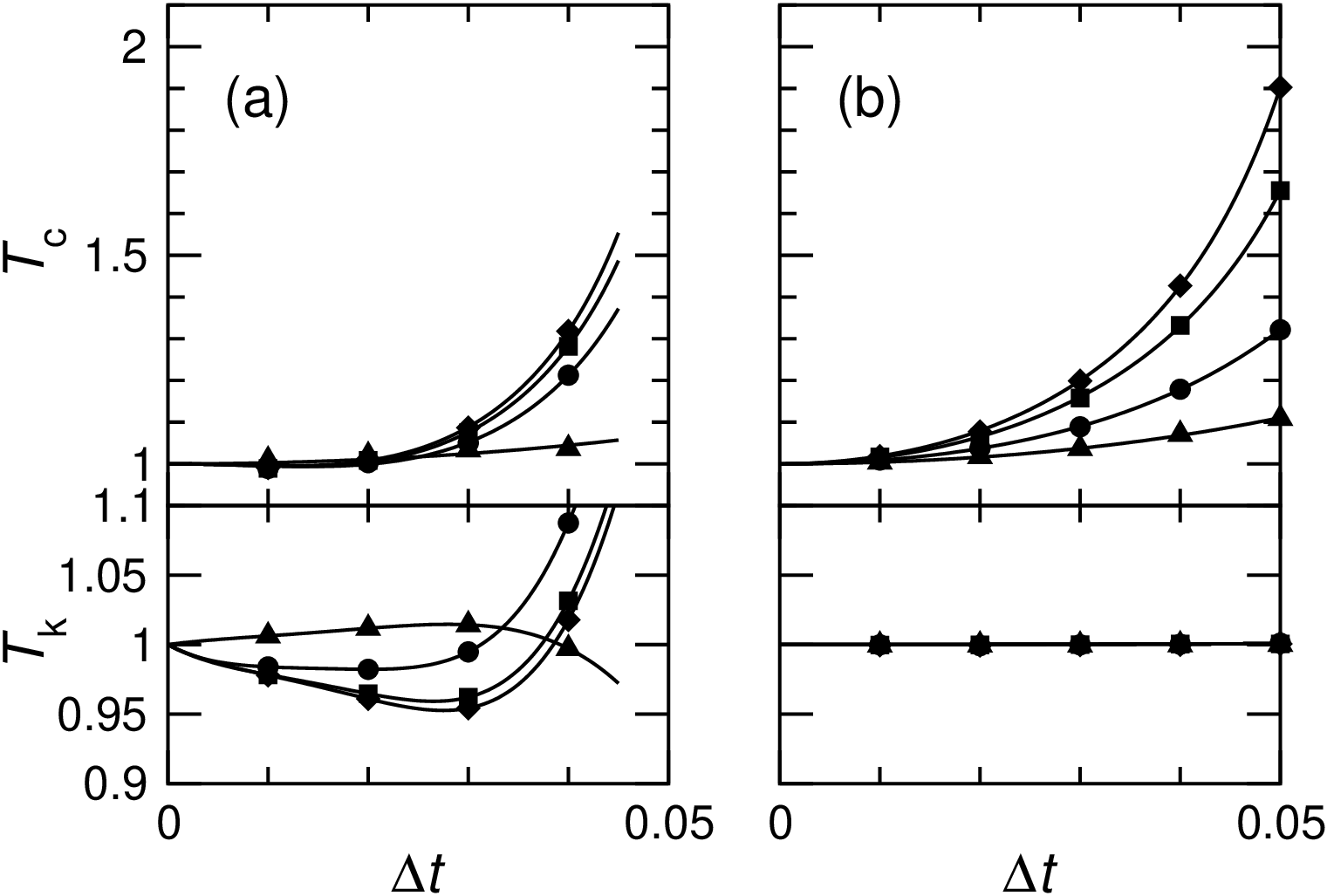}} \hrulefill
\end{figure}

For the membrane simulations, temperatures for each of the bead types (species)
are reported in Figs~\ref{fig:3} and \ref{fig:4}.  There are several points of
interest. Firstly, the kinetic temperatures for different species are different,
as already noted \citep{jakobsen.af:2005.a}. The discrepancies between
$T_\text{k}$ and $T$ for the lipid beads are somewhat worse than in the case of
simple water at the same timesteps, of order a few percent. The configurational
temperatures are also different for the different species. For the lipid beads,
they are all in \emph{much} poorer agreement with the desired temperature, than
for the water: at $\Delta t=0.05$, $T_\text{c}$ is too high by 80--90\% for some
of the tail beads. This indicates a serious departure from equilibrium.
Secondly, numbering the tail beads $\text{T}_1$--$\text{T}_6$ with $\text{T}_1$
next to the head bead, the results for both $T_\text{k}$ and $T_\text{c}$ divide
naturally into groups: $\{\text{H},\text{T}_6\}$, $\{\text{T}_1,\text{T}_5\}$,
$\{\text{T}_2,\text{T}_3,\text{T}_4\}$, and $\{\text{W}\}$. Within each group,
the temperatures agree, almost to within the statistical error bars, and no
attempt is made to distinguish them in the figures. This
suggests \citep{jakobsen.af:2005.a} that the intramolecular stretching and bending
potential terms (which are the same within each group) are the dominant factors
in the problems of maintaining equilibrium; in the following section, some
support will be given to this view, but it is also argued that the
intermolecular potentials make a contribution. Of the four algorithms studied,
the DPD, Peters, and heavily-thermostatted Lowe method perform in a similar way
to each other. The behaviour of the Stoyanov algorithm is unusual: although
$T_\text{c}$ is quite close to $T$ up to $\Delta t=0.03$, $T_\text{k}$ shows an
instability. As $\Delta t$ increases, the water temperature falls dramatically
below $T$, while the other species become much too hot. This effect makes the
simulation unphysical for $\Delta t=0.05$.

\begin{figure}[tbp]
\hrulefill
\caption{\label{fig:5}%
  Density and temperature profiles in the bilayer region for membrane
  simulations using the Peters thermostat. Top two panels show configurational
  and kinetic temperatures as functions of timestep. Solid lines: $\Delta
  t=0.05$. Dashed lines: $\Delta
  t=0.03$. Dot-dashed lines: $\Delta
  t=0.01$. Bottom panel shows, for reference, the local density, measured at $\Delta
  t=0.01$. Solid lines: H beads. Long dashes: T
  beads. Short dashes: W beads.}
\centerline{\includegraphics[width=0.6\textwidth,clip]{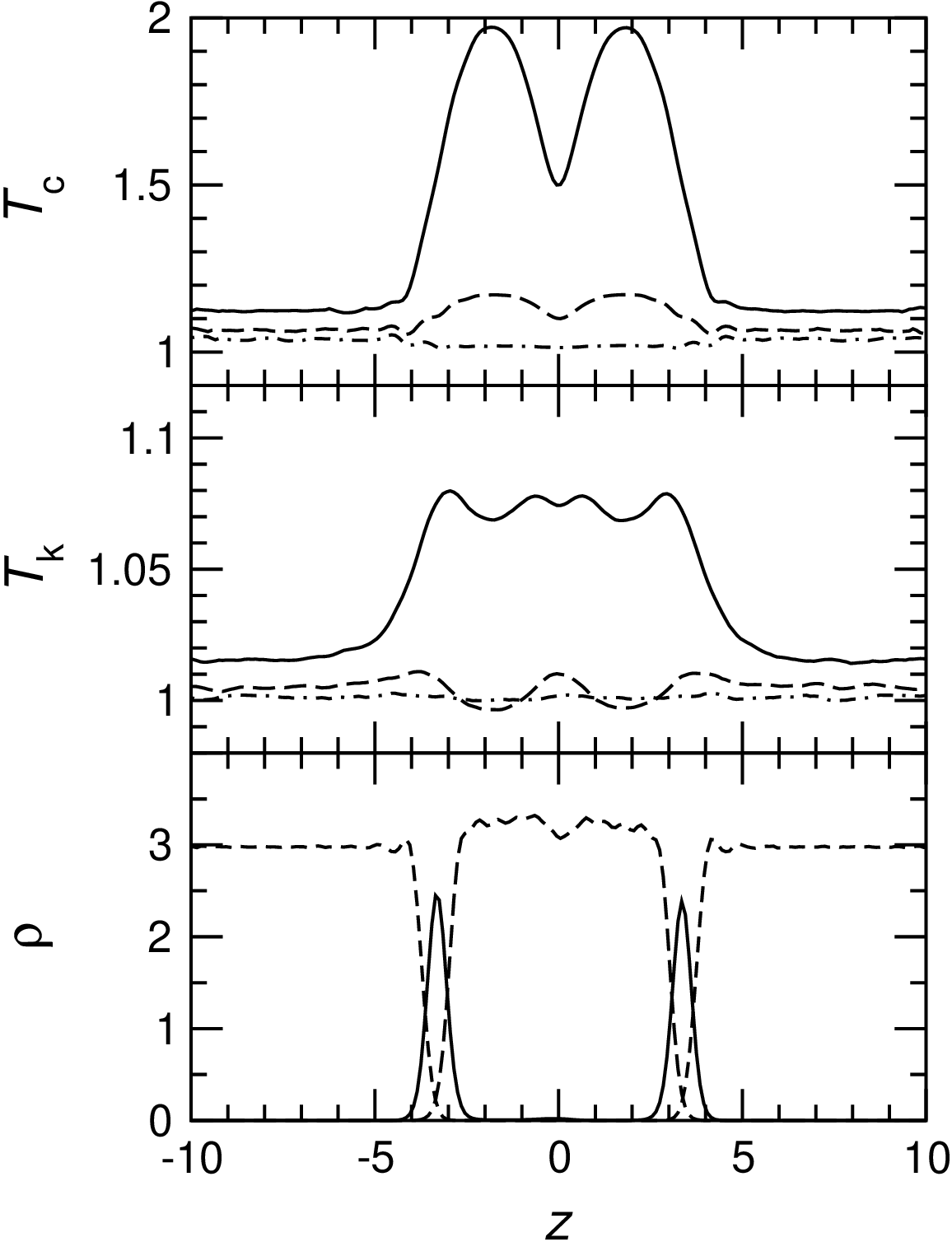}} \hrulefill
\end{figure}
\begin{figure}[tbp]
\hrulefill
\caption{\label{fig:6}%
  Density and temperature profiles in the bilayer region for membrane
  simulations at $\Delta t=0.03$ using the different thermostats. Top two panels
  show configurational and kinetic temperatures. Solid lines: Lowe thermostat
  with $\Gamma=1/\Delta t$.  Long dashes: Peters thermostat with $\sigma=3$.
  Short dashes: DPD algorithm with $\sigma=3$.  Dash-dotted line: Stoyanov-Groot
  thermostat with $\Gamma=0$. Bottom panel shows, for reference, the local
  density, measured at $\Delta t=0.01$. Solid lines: H beads. Long dashes: T
  beads. Short dashes: W beads.}
\centerline{\includegraphics[width=0.6\textwidth,clip]{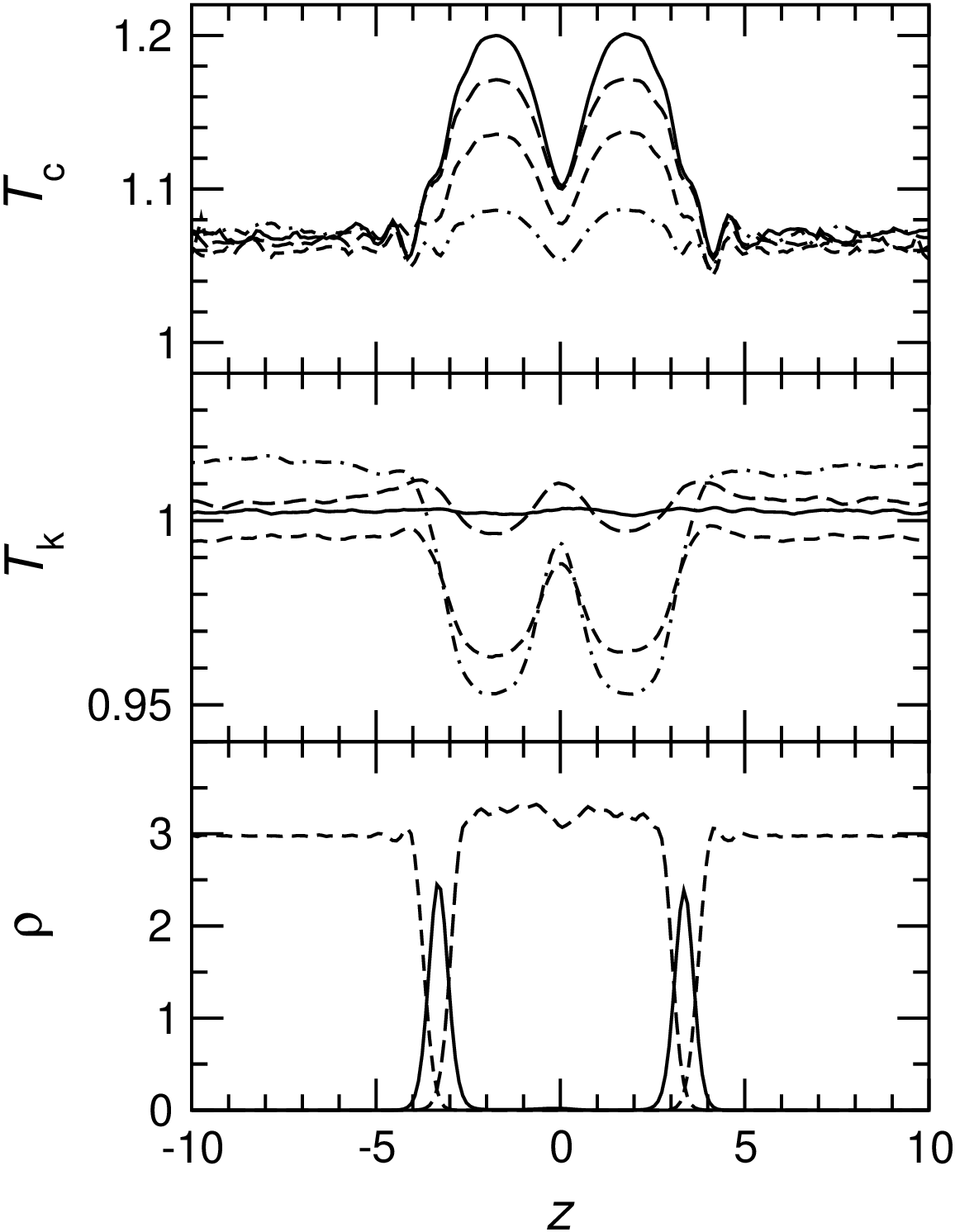}} \hrulefill
\end{figure}

Density and temperature profiles are shown in Fig.~\ref{fig:5}, selecting the
Peters thermostat as being representative. \citet{jakobsen.af:2005.a} report a
time step dependence of the density profiles, specifically a change in the
(small) concentrations of head group beads in the interior of the bilayer. The
present simulations do not confirm such a systematic effect; however, the
transverse dimensions of the system are much smaller than those of
Ref.~\citep{jakobsen.af:2005.a}, so the relevant statistics and timescales may
prevent resolution of the effect here. The temperature profiles are as expected
from Figs~\ref{fig:3} and \ref{fig:4}, with the most dramatic indicators of lack
of equilibrium in the regions occupied by tail beads
$\{\text{T}_2,\text{T}_3,\text{T}_4\}$.  A comparison of the different
algorithms at fixed timestep $\Delta t=0.03$ is given in Fig~\ref{fig:6}. In the
lipid tail region the Lowe thermostat (with the parameter $\Gamma$ set to the
maximum allowed value) gives the worst $T_\text{c}$, about 20\% too high, while
the Stoyanov thermostat (with $\Gamma=0$) gives the best, about 8\% too high;
however the performance order is reversed for $T_\text{k}$ with the Lowe
thermostat (by construction) being perfectly accurate and the Stoyanov
thermostat being overdamped by about 5\%. The behaviour in the water region does
not give a good indicator of the lack of equilibrium within the membrane.

\begin{figure}[tbp]
\hrulefill
\caption{\label{fig:7}%
  Density and pressure profiles in the bilayer region for membrane simulations
  using the Peters thermostat. Top panel: normal component $P_\text{N}$ as a
  function of timestep: Solid lines: $\Delta t=0.05$. Dashed lines: $\Delta
  t=0.03$. Dot-dashed lines: $\Delta t=0.01$. Middle panel: surface tension
  integrand, $P_\text{N}-P_\text{T}$, measured at $\Delta t=0.01$. Bottom panel:
  local density, measured at $\Delta t=0.01$. Solid lines: H beads. Long dashes:
  T beads. Short dashes: W beads.}
\centerline{\includegraphics[width=0.6\textwidth,clip]{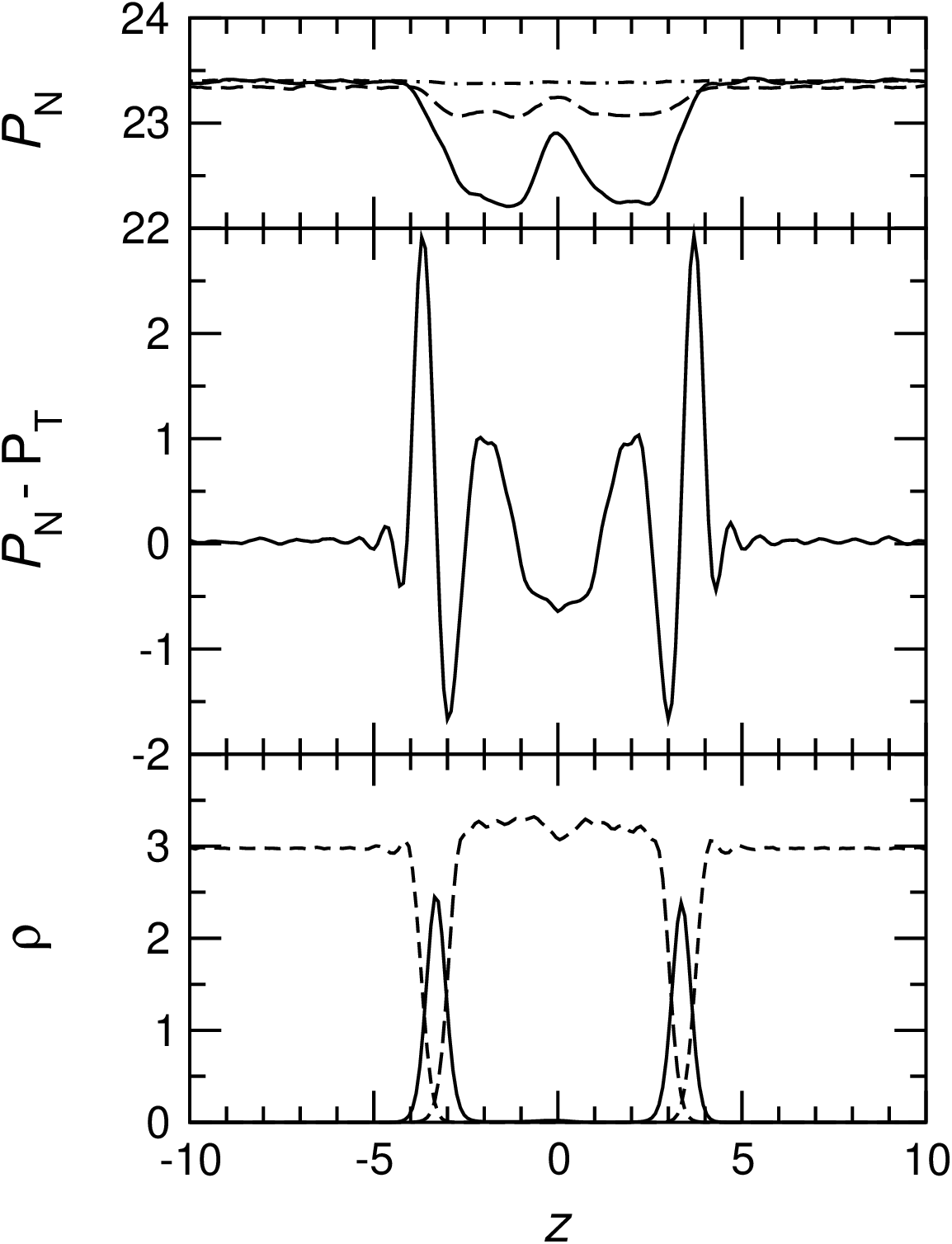}} \hrulefill
\end{figure}

It is of interest to compare these profiles with the pressure tensor results
which were also previously found to exhibit an anomaly \citep{jakobsen.af:2005.a}.
The pressure profiles are calculated in the Irving-Kirkwood
convention \citep{schofield.p:1982.a} using a method essentially the same as that
described by \citet{goetz.r:1998.a}.  The results for the
Peters thermostat are shown in Fig.~\ref{fig:7}. The component $P_\text{N}$
normal to the bilayer is not constant, contrary to what must hold for a system
at equilibrium \citep{jakobsen.af:2005.a}.  For $\Delta t=0.05$ the error is
comparable in magnitude to the physically interesting quantity, the difference
$P_\text{N}-P_\text{T}$ between normal and transverse components, whose integral
gives the surface tension. There is some cancellation of errors, since the
anomaly appears in both components \citep{jakobsen.af:2005.a}; however, such an
error is clearly undesirable. The point to be made here is that the
configurational temperature profiles of Fig.~\ref{fig:5} are echoed in the
pressure profiles of Fig.~\ref{fig:7}.

\section{Analysis and Discussion}
It is evident that the configurational temperature gives a different
perspective, from the kinetic temperature, on simulation equilibrium, even for
simple fluids. Also the intramolecular bond stretching and bending potentials
used in membrane simulations have a direct effect on the stability of the
algorithm, which can be estimated via $T_\text{c}$, amongst other indicators.

Some insight into the information contained in $T_\text{c}$ may be obtained by
considering a very simple model: the 1-D harmonic oscillator, with hamiltonian
\begin{equation*}
\mathcal{H} = p^2/2m + \tfrac{1}{2}k x^2 \equiv K(p) + U(x) \:.
\end{equation*}
It is known \citep{toxvaerd.s:1994.a,gans.j:2000.a} that the velocity Verlet
algorithm conserves \emph{exactly} a ``shadow hamiltonian'' which we may write
in the form
\begin{align*}
\mathcal{H}^\ddag(x,p) &= p^2/2m +
\tfrac{1}{2}k x^2 \bigl(1-\phi\bigr) \equiv K(p) + U^\ddag(x)
\\
\text{where}\qquad \phi &=\tfrac{1}{4}(k/m) \Delta t^2 \;.
\end{align*}
A dynamical algorithm which alternates periods
of velocity Verlet with velocity reselection from the canonical ensemble at
temperature $T$ (analogous to the ``Lowe'' algorithm of DPD) will actually
generate
momenta from the correct Maxwell-Boltzmann distribution
$\exp(-\mathcal{K}/k_\text{B}T)$
and
configurations from the ``shadow distribution''
$\exp(-\mathcal{U}^\ddag/k_\text{B}T)$.  An immediate consequence is
\begin{equation*}
\tfrac{1}{2}k_\text{B}T = \left\langle p^2/2m \right\rangle =
\left\langle \tfrac{1}{2}k x^2\bigl(1-\phi\bigr) \right\rangle \;.
\end{equation*}
Since the force $f=-kx$ and $\partial^2\mathcal{H}/\partial x^2=k$, the kinetic and
configurational temperatures are
\begin{subequations}
\begin{align}
k_\text{B}T_\text{k} &= \left\langle  p^2/m\right\rangle = k_\text{B}T
\\
k_\text{B}T_\text{c} &= 
\frac{\bigl\langle f^2\bigr\rangle}{\bigl\langle\partial^2\mathcal{H}/\partial
  x^2\bigr\rangle} 
= \frac{k_\text{B}T}{1-\phi} \;.
\label{eqn:shoTc}
\end{align}
\end{subequations}
For a 3D isotropic harmonic oscillator $U=\frac{1}{2}k |\vec{r}|^2$,
$\vec{f}=-k\vec{r}$ the same formula applies, with $k=\frac{1}{3}\nabla^2 U$,
and it is straightforward to extend it to the anisotropic case.  The effect of
an over-large time step is to reduce the effective force constant dictating the
configurational distribution, increase the mean squared displacement, and (in
proportion) increase the mean squared force which gives the value of
$T_\text{c}$. The above equation is a reasonable first approximation to the
curves shown in Figs~\ref{fig:1}, \ref{fig:2}. For the lipids, assuming that the
intramolecular bond potentials are approximately harmonic, a given fractional
error in $T_\text{c}$ for each bead may approximately translate into a
corresponding fractional error in mean-squared displacement which, when
compounded along the chain, becomes a more significant error in overall
flexibility.

It is tempting to try to estimate these contributions to $T_\text{c}$, for the
isolated lipid chain. In the harmonic approximation, the appropriate force
constants may be evaluated, taking the lipids to be in their minimum energy
configuration, and counting all the bonds in which a bead participates.  In
Table~\ref{tab:1} the sum (over $x$, $y$ and $z$ directions) of these predicted
force constants (for $k_\text{stretch}=128$, $\ell_0=0.5$, $k_\text{bend}=20$)
are compared with measured average Laplacians from simulations of a single chain
at $T=1$, and simulations of the full fluid system at $T=1$, both conducted at
short timesteps.
\begin{table}[tbp]
\hrulefill
\caption{\label{tab:1}%
Laplacian $\nabla_i^2 U$ for beads in lipid chain calculated for isolated
chain at $T=0$, as an equilibrium simulation average at $T=1$ for an isolated
chain, from the full condensed phase simulation at $T=1$, and from a fit to
the measured $T_\text{c}$, Fig.~\ref{fig:4}b, assuming a locally isotropic
harmonic potential. 
Figures in parentheses represent the estimated error in the last digit.}
\begin{center}
\begin{tabular}{|l|rr|rr|}\hline
 & \multicolumn{2}{c|}{single chain} & \multicolumn{2}{c|}{condensed phase} \\ 
Bead & $T=0$ & $T=1$ & $T=1$ & Fit  \\ \hline
H            & 288  & 273(2) & 491(1) & 1157(1) \\
$\text{T}_1$ & 1056 & 853(6) & 1132(1) & 1893(1) \\
$\text{T}_2$ & 1216 & 979(5) & 1272(1) & 2272(1) \\
$\text{T}_3$ & 1216 & 986(7) & 1266(2) & 2327(1) \\
$\text{T}_4$ & 1216 & 972(8) & 1264(2) & 2277(1) \\
$\text{T}_5$ & 1056 & 865(8) & 1116(2) & 1884(1) \\
$\text{T}_6$ & 288  & 267(3) & 438(1) & 1123(1) \\
W            &     & & 122.8(1) & 479(1) \\ \hline
\end{tabular}
\end{center}
\hrulefill
\end{table}
The figures partly explain the close similarity in configurational temperatures
within the groups of beads mentioned in the previous section, in terms of
intramolecular potentials. There is some reduction in the average Laplacians due
to fluctuations of the single chain at $T=1$ compared with $T=0$. However, in
the full condensed phase simulation, this is more than compensated by an
intermolecular contribution of order 200--250 in these reduced units, so part of
this coincidence is due to the similar environments, and identical repulsion
parameters between like beads, for this particular model.

The simulated fluids are not actually harmonic, and so the above analysis should
not be over-interpreted. The final column of Table~\ref{tab:1} shows the results
of fitting eqn~\eqref{eqn:shoTc} to the Lowe thermostat results shown in
Fig.~\ref{fig:4}b. The fits are excellent, but the fitted parameters are all
significantly larger than would have been expected. Partly, this can be
explained by the anisotropy of the local harmonic potential: the stiffest
potentials will dominate the breakdown of the algorithm. However one should also
remember that the real shadow hamiltonian is more complicated than a simple
harmonic potential, and will not split in the same way into a kinetic part plus
a renormalized configurational part.

Some effort has been made to develop practical methods to estimate the shadow
Hamiltonian for complex systems \citep{skeel.rd:2001.a,engle.rd:2005.a}. Knowledge
of this function may help not only the monitoring of algorithm performance, but
also the implementation of more efficient algorithms such as hybrid Monte
Carlo \citep{izaguirre.ja:2004.a}. This method differs from DPD in the rejection
of moves so as to guarantee that the desired ensemble is sampled. Returning to
the points made by \citet{hafskjold.b:2004.a}, it is
notable that \emph{smoothness} of the potential is a highly desirable feature in
the accurate construction of shadow hamiltonians. To illustrate this,
\citet{engle.rd:2005.a} considered a simple harmonic oscillator
potential, modified by splitting into two halves at the minimum, with a flat
piece inserted between them. The consequent lack of smoothness was shown to
degrade conservation of the shadow hamiltonian. Such a potential is a possible
one-dimensional model system for the numerical aspects of DPD dynamics: the
oscillation back and forth represents the motion of a DPD bead out of the
repulsive quadratic potential of a neighbour, and into the repulsive range of a
different neighbour. In the absence of an improved algorithm to handle the DPD
conservative forces, using a shorter timestep for them seems the best
approach \citep{peters.eajf:2004.a}.

\section{Conclusions}
The essential conclusion of this paper is that stiffer potentials, and stronger
conservative forces, require shorter timesteps. This is hardly a new point, and
is increasingly being recognised in the DPD
community \citep{peters.eajf:2004.a,hafskjold.b:2004.a,jakobsen.af:2005.a}, where
the principal danger is that the effects of an over-long timestep may be
disguised by the thermostat.  It is recommended here that the configurational
temperature $T_\text{c}$ be measured and reported in DPD simulations, along with
other indicators that the system is not at equilibrium, so as to address this
danger.  $T_\text{c}$ is easy to calculate, and appears to be sensitive to some
of the artifacts that appear in other properties such as the pressure tensor
profile in membranes.

The degree of accuracy of sampling required will, no doubt, vary from one
application to another. For the standard water model, \citet{groot.rd:1997.a}
recommended timesteps $\Delta t \approx 0.05$, so as to achieve an accuracy of
order 1\% in $T_\text{k}$. Applying the same criterion to $T_\text{c}$ would
lead to a revised recommendation, $\Delta t \approx 0.015$. Modern thermostats
are dramatically better at controlling $T_\text{k}$, and give modest
improvements in $T_\text{c}$. \citet{stoyanov.sd:2005.a} note,
for instance, that their new method keeps $T_\text{k}$ within the desired 1\%
range even for $\Delta t=0.09$; however, such a timestep would result in an
error in $T_\text{c}$ between 24\% and 44\% depending on $\Gamma$. Also, for
multicomponent systems such as lipid membranes, there may be substantial
imbalances in the distribution of $T_\text{k}$ and $T_\text{c}$ between the
different species. Such large discrepancies indicate lack of equilibrium, i.e.\ 
configurations are being sampled incorrectly.

It would, no doubt, be possible to devise thermostats that control the value of
$T_\text{c}$ as well as $T_\text{k}$, and that act separately on the different
types of bead. In such a case, it becomes important to measure other,
independent, quantities as a function of time step, and to bear in mind that no
single quantity acts as an indicator of equilibrium. 
\section*{Acknowledgments}
The simulations reported here were performed on the computing facilities of the
Centre for Scientific Computing, University of Warwick. Funding was provided by
EPSRC Materials Modelling Initiative grant GR/S80127.
\section*{Appendix}
The calculation of the Laplacian for each of the potential terms used in this
study proceeds as follows. The DPD repulsive potential between $i$ and $j$, with
$\vec{r}_{ij}=\vec{r}_i-\vec{r}_j$, takes the form (assuming $r_{ij} \leq r_\text{c}$)
\begin{equation*}
u = \frac{1}{2} \left(\frac{\alpha}{r_\text{c}}\right)
\bigl(r_{ij}-r_\text{c}\bigr)^2 \;.
\end{equation*}
The gradients (the negatives of which are the conservative forces of eqn~\eqref{eqn:conforcea}) are
\begin{equation*}
\vec{\nabla}_i u =-\vec{\nabla}_j u 
= \alpha \left(\frac{r_{ij}}{r_\text{c}}-1\right)\hat{\vec{r}}_{ij} \;,
\end{equation*}
and the Laplacians are
\begin{equation*}
\nabla_i^2 u = \nabla_j^2 u  = \alpha \left(
  \frac{3}{r_\text{c}}-\frac{2}{r_{ij}} \right) \;.
\end{equation*}
The formulae for the bond stretching potential of eqn~\eqref{eqn:potstretch} are
identical, with the replacements $r_\text{c}\rightarrow\ell_0$ and
$\alpha\rightarrow k_\text{stretch}\ell_0$.

For the angle-bending potential of eqn~\eqref{eqn:potbend} between successive beads
$i$, $j$, $k$, define $c_{ij}=\vec{r}_{ij}\cdot\vec{r}_{ij}$,
$c_{jk}=\vec{r}_{jk}\cdot\vec{r}_{jk}$, $c_{ijk}=\vec{r}_{ij}\cdot\vec{r}_{jk}$.
The gradients of these scalar products with respect to individual positions are
\begin{align*}
\vec{\nabla}_i c_{ij} &= 2\vec{r}_{ij}
&
\vec{\nabla}_i c_{jk} &= 0
&
\vec{\nabla}_i c_{ijk} &= \vec{r}_{jk}
\\
\vec{\nabla}_j c_{ij} &= -2\vec{r}_{ij}
&
\vec{\nabla}_j c_{jk} &= 2\vec{r}_{jk}
&
\vec{\nabla}_j c_{ijk} &= \vec{r}_{ij}-\vec{r}_{jk}
\\
\vec{\nabla}_k c_{ij} &= 0
&
\vec{\nabla}_k c_{jk} &= -2\vec{r}_{jk}
&
\vec{\nabla}_k c_{ijk} &= -\vec{r}_{ij} \;.
\end{align*}
The Laplacians are
\begin{align*}
\nabla_i^2 c_{ij} &= 6
&
\nabla_i^2 c_{jk} &= 0
&
\nabla_i^2 c_{ijk} &= 0
\\
\nabla_j^2 c_{ij} &= 6
&
\nabla_j^2 c_{jk} &= 6
&
\nabla_j^2 c_{ijk} &= -6
\\
\nabla_k^2 c_{ij} &= 0
&
\nabla_k^2 c_{jk} &= 6
&
\nabla_k^2 c_{ijk} &= 0 \;.
\end{align*}
The potential has the form $u_\text{bend}=k_\text{bend}(1-C)$, where
\begin{equation*}
C = \cos\theta_{ijk} = c_{ij}^{-1/2} \, c_{jk}^{-1/2} \, c_{ijk} \;.
\end{equation*}
The necessary gradients are
\begin{align*}
\vec{\nabla}_i C &= 
\bigl(\vec{\nabla}_i c_{ij}^{-1/2} \bigr) c_{jk}^{-1/2} c_{ijk}
+
c_{ij}^{-1/2} \bigl(\vec{\nabla}_i c_{jk}^{-1/2}\bigr)  c_{ijk}
+
c_{ij}^{-1/2} c_{jk}^{-1/2} \bigl(\vec{\nabla}_i  c_{ijk}\bigr) 
\\ &= c_{ij}^{-1/2} c_{jk}^{-1/2} 
\left( \vec{r}_{jk} - \frac{c_{ijk}}{c_{ij}}\vec{r}_{ij}\right)
\\
\vec{\nabla}_j C &= 
\bigl(\vec{\nabla}_j c_{ij}^{-1/2} \bigr) c_{jk}^{-1/2} c_{ijk}
+
c_{ij}^{-1/2} \bigl(\vec{\nabla}_j c_{jk}^{-1/2}\bigr)  c_{ijk}
+
c_{ij}^{-1/2} c_{jk}^{-1/2} \bigl(\vec{\nabla}_j  c_{ijk}\bigr) 
\\ &= c_{ij}^{-1/2} c_{jk}^{-1/2} 
\left( \Bigl(1+\frac{c_{ijk}}{c_{ij}}\Bigr)\vec{r}_{ij} 
-\Bigl(1+\frac{c_{ijk}}{c_{jk}}\Bigr)\vec{r}_{jk}\right)
\\
\vec{\nabla}_k C &= 
\bigl(\vec{\nabla}_k c_{ij}^{-1/2} \bigr) c_{jk}^{-1/2} c_{ijk}
+
c_{ij}^{-1/2} \bigl(\vec{\nabla}_k c_{jk}^{-1/2}\bigr)  c_{ijk}
+
c_{ij}^{-1/2} c_{jk}^{-1/2} \bigl(\vec{\nabla}_k  c_{ijk}\bigr) 
\\ &= c_{ij}^{-1/2} c_{jk}^{-1/2} 
\left( \frac{c_{ijk}}{c_{jk}}\vec{r}_{jk}-\vec{r}_{ij}\right) \;.
\end{align*}
The Laplacians are given by
\begin{align*}
\nabla_i^2 C &= 
\bigl(\nabla_i^2 c_{ij}^{-1/2} \bigr) c_{jk}^{-1/2} c_{ijk}
+
c_{ij}^{-1/2} \bigl(\nabla_i^2 c_{jk}^{-1/2}\bigr)  c_{ijk}
+
c_{ij}^{-1/2} c_{jk}^{-1/2} \bigl(\nabla_i^2  c_{ijk}\bigr) 
\\ &
+2 c_{ij}^{-1/2} 
\bigl(\vec{\nabla}_i c_{jk}^{-1/2}\cdot\vec{\nabla}_i c_{ijk}\bigr)
+2 c_{jk}^{-1/2}
\bigl(\vec{\nabla}_i c_{ijk}\cdot\vec{\nabla}_i c_{ij}^{-1/2} \bigr)
+2 c_{ijk}
\bigl(\vec{\nabla}_i c_{ij}^{-1/2} \cdot\vec{\nabla}_i c_{jk}^{-1/2}\bigr)
\\
&= -2 c_{ij}^{-3/2} c_{jk}^{-1/2} c_{ijk}
\\
\nabla_j^2 C &= 
\bigl(\nabla_j^2 c_{ij}^{-1/2} \bigr) c_{jk}^{-1/2} c_{ijk}
+
c_{ij}^{-1/2} \bigl(\nabla_j^2 c_{jk}^{-1/2}\bigr)  c_{ijk}
+
c_{ij}^{-1/2} c_{jk}^{-1/2} \bigl(\nabla_j^2  c_{ijk}\bigr) 
\\ &
+2 c_{ij}^{-1/2} 
\bigl(\vec{\nabla}_j c_{jk}^{-1/2}\cdot\vec{\nabla}_j c_{ijk}\bigr)
+2 c_{jk}^{-1/2}
\bigl(\vec{\nabla}_j c_{ijk}\cdot\vec{\nabla}_j c_{ij}^{-1/2} \bigr)
+2 c_{ijk}
\bigl(\vec{\nabla}_j c_{ij}^{-1/2} \cdot\vec{\nabla}_j c_{jk}^{-1/2}\bigr)
\\
&= -2 c_{ij}^{-3/2} c_{jk}^{-3/2} \bigl(
c_{ijk}^2 + c_{ijk} (c_{ij}+ c_{jk}) + c_{ij}c_{jk}
\bigr)
\\
\nabla_k^2 C &= 
\bigl(\nabla_k^2 c_{ij}^{-1/2} \bigr) c_{jk}^{-1/2} c_{ijk}
+
c_{ij}^{-1/2} \bigl(\nabla_k^2 c_{jk}^{-1/2}\bigr)  c_{ijk}
+
c_{ij}^{-1/2} c_{jk}^{-1/2} \bigl(\nabla_k^2  c_{ijk}\bigr) 
\\ &
+2 c_{ij}^{-1/2} 
\bigl(\vec{\nabla}_k c_{jk}^{-1/2}\cdot\vec{\nabla}_k c_{ijk}\bigr)
+2 c_{jk}^{-1/2}
\bigl(\vec{\nabla}_k c_{ijk}\cdot\vec{\nabla}_k c_{ij}^{-1/2} \bigr)
+2 c_{ijk}
\bigl(\vec{\nabla}_k c_{ij}^{-1/2} \cdot\vec{\nabla}_k c_{jk}^{-1/2}\bigr)
\\
&= -2 c_{ij}^{-1/2} c_{jk}^{-3/2} c_{ijk} \;.
\end{align*}
All of the above results should be multiplied by $(-k_\text{bend})$ to give
the gradients and Laplacians for the bending potential of
eqn~\eqref{eqn:potbend}.

It is worth emphasizing that the dissipative and random terms in the DPD
equations \eqref{eqn:dpd} play no role in the calculation of Laplacians and squared forces which
go into the configurational temperature, eqn~\eqref{eqn:contemp}.
\bibliography{journals,main}
\end{document}